# Comment on "On the Extraction of Purely Motor EEG Neural Correlates during an Upper Limb Visuomotor Task"

Patrick Ofner[1,2], Joana Pereira[2,3], Reinmar Kobler[4,2], Andreas Schwarz[5] and Gernot R. Müller-Putz[2,6]

[1] Bernstein Center Freiburg, University of Freiburg, Freiburg, Germany
[2] Institute of Neural Engineering, Graz University of Technology, Graz, Austria
[3] Stereotactic and Functional Neurosurgery Department, University Medical Center Freiburg, Freiburg, Germany
[4] RIKEN Center for Advanced Intelligence Project, Kyoto, Japan
[5] Section Experimental Neurorehabilitation, Spinal Cord Injury Center, Heidelberg University Hospital, Heidelberg, Germany
[6] BioTechMed Graz, Graz, Austria

Correspondence should be addressed to P.O. (email: patrick@ofner.science)

# Abstract

Bibian et al. show in their recent paper (Bibián et al. 2021) that eye and head movements can affect the EEG-based classification in a reaching motor task. These movements can generate artefacts that can cause an overoptimistic estimation of the classification accuracy. They speculate that such artefacts jeopardise the interpretation of the results from several motor decoding studies including our study (Ofner et al. 2017). While we endorse their warning about artefacts in general, we do have doubts whether their work supports such a statement with respect to our study. We provide in this commentary a more nuanced contextualization of our work presented in Ofner et al. and the type of artefacts investigated in Bibian et al.

# Introduction

Bibian et al. (Bibián et al. 2021) demonstrate the influence of artefacts on the movement decoding from EEG signals recorded during an upper limb visuomotor task. The study participants performed a reaching task in the horizontal plane with their right hand under two conditions: constrained and unconstrained. In the unconstrained condition, the study authors did not instruct participants to reduce eye or head movements, whereas, in the constrained condition, participants were explicitly instructed to avoid eye and head movements. Using a linear support vector machine (SVM), the authors classified 4 reaching directions from EEG low-frequency time-domain (LFTD) features as well as power spectral features. The authors found for LFTD features a significantly higher classification accuracy for the unconstrained condition compared to the constrained condition. Moreover, the performance of the LFTD-based classification in the constrained condition was around chance-level. The authors reasoned that likely task-correlated artefacts contaminated the EEG signals, and thus allowed to successfully classify the reaching directions from LFTD features in the unconstrained condition. Based on this reasoning, the authors questioned the validity of the results obtained in previous upper-limb movement decoding studies based on low-frequency EEG signals. They speculate that movement decoding in those studies may considerably depend on artefacts.

In particular, the authors refer to our study (Ofner et al. 2017) and furthermore claim that (Ofner et al. 2017) *"identified EEG correlates to identify different movements from the same arm using a protocol that was not designed to decorrelate the task and its associated-task artefacts, thus not excluding the possibility that artefacts had an influence on the decoding accuracy"*.

## Clarification on the Interpretation of our Work

We see problems in their interpretation of our work. Before we detail our concerns, we would like to emphasise that we are not questioning the results provided in (Bibián et al. 2021) or their warning of artefacts. We do recognize the importance of careful paradigm design and minimization of eye and head-movement related artefacts which can potentially contaminate low-frequency EEG signals (Philips et al. 2014; Kobler, Sburlea, Lopes-Dias, et al. 2020).

We would like to put (Bibián et al. 2021) and our work in context. We have studied a different motor task than (Bibián et al. 2021). In our work in Ofner et al., participants were asked to perform single movements (forearm pronation, forearm supination, elbow flexion, elbow extension, palmar grasp, hand opening) rather than complex reaching movements. Subsequently, we classified which movement was performed. Importantly, our study protocol did not involve spatially separated reach targets. Thus, study participants were not required to make task-associated eye or head movements to spatial targets like in the unconstrained condition in (Bibián et al. 2021). We furthermore instructed the participants to keep their gaze on a fixation cross on a computer screen so that head and eye movements were minimised. In contrast to Bibian et al., we did not employ a visuomotor task requiring eye-hand coordination. Thus, our employed study protocol was more restrictive than the constrained condition of (Bibián et al. 2021). The authors' suggestion that our decoding results could be considerably affected by artefacts is therefore not comprehensible. Furthermore, their claim that our protocol *"was not designed to decorrelate the task and its associated-task artefacts"* is in this context not appropriate as the task-associated artefacts investigated in Bibian et al. (i.e. eye and head movement-related artefacts) were simply not provoked by our study protocol. Apart from eye and head movement-related artefacts, arm/hand movements themselves can potentially introduce artefacts in LFTD features that covary with the experimental variable (Philips et al. 2014). Their contribution to the LFTD features in Bibian et al. was apparently negligible, as the movement direction could not be inferred in the constrained condition. We did investigate whether our study results were affected by arm/hand movement-related artefacts. For this purpose, we have provided source imaging results in (Ofner et al. 2017) which indicate that executed movements were indeed decoded from motor-related areas. In this context, we would also like to point out that we have later successfully decoded attempted single movements from persons with spinal cord injury without overt movements where neither head, eye nor hand movements were associated with the task (Ofner et al. 2019). We conclude therefore that the authors in (Bibián et al. 2021) do not provide evidence that artefacts jeopardise the interpretation of the results presented in (Ofner et al. 2017).

# On the Decoding of Movements

Finally, we would like to briefly discuss three aspects that we think can impact the decoding performance, and which could explain why the authors could not decode movement targets/directions in the constrained condition.

Data alignment and feature extraction window
Wang et al. showed in a magnetoencephalography study that delaying the movement in a center-out task with an explicit go cue has an adverse effect on the direction decoding performance (Wang et al. 2010). We have reported in (Kobler, Kolesnichenko, et al. 2020) that the movement direction can be better decoded from direction cue-aligned EEG data than from movement onset-aligned EEG data. This highlights that especially brain processes related to the processing of the direction cue provide decodable information about movement direction. The employed go cue in (Bibián et al. 2021) as well as the time-locking to the movement onset may have therefore impeded the direction decoding. Furthermore, as their feature window started at least two seconds after the presentation of the direction cue, it might not have contained sufficient decodable information anymore provided by the direction cue. Thus, while the employed go cue is a choice of the paradigm design and therefore sometimes unavoidable, the time-alignment and the position of the feature window might not have been ideal for direction/target classification.

Data preprocessing
In (Bibián et al. 2021), common average reference (CAR) filtering was applied *"considering only the electrodes placed over the sensorimotor cortex"*. However, such processing can subtract in the case of low-frequency EEG signals the movement direction signal from the EEG. This is because particularly low-frequency EEG signals are widespread due to volume conduction and the low-pass properties of the tissue/networks (Buzsáki, Anastassiou, and Koch 2012). Thus, a CAR filter considering only electrodes in the vicinity of the signal source(s) could decrease the signal-to-noise ratio. In line with this, the findings of Garipelli et al. (Garipelli, Chavarriaga, and Millán 2013) suggest that a standard CAR filter (i.e., all electrodes) is more suitable than focal spatial filters to extract LFTD signals. However, Jochumsen et al. (Jochumsen et al. 2015) found that a large Laplacian spatial filter is better suited than CAR to extract discriminative features from movement-related cortical potentials. A reason for this discrepancy may be that Jochumsen et al. included higher frequency components (up to 10 Hz) in which noise can potentially be more effectively filtered out with focal spatial filters.

Classification

The adversely low trial-to-feature ratio in (Bibián et al. 2021) could be challenging for an SVM classifier given that the authors do not report any regularisation measure (50 trials per class vs. 100 features). Here, the inclusion of a regularisation term in the optimization objective may have been beneficial. Alternatively, a shrinkage linear discriminant analysis classifier could have been used, since it can typically handle LFTD signals with low trial-to-feature ratios (Blankertz et al. 2011).

We do not question the results in (Bibián et al. 2021) or have any doubts about their conclusion that artefact contamination is a crucial aspect when dealing with motor tasks and their discrimination based on brain signals, but the aforementioned aspects could perhaps explain why no classification was possible in their constrained condition. We further want to emphasise that the authors might well have considered these methodological changes but did not find them beneficial for decoding in their case.

Finally, for future visuomotor studies, we would like to point out recent progress made by our group in attenuating eye-related artefacts. The approach introduced in (Kobler, Sburlea, Lopes-Dias, et al. 2020) attenuates covarying eye movement and blink artefacts to noise level, while retaining kinematic information (Kobler, Kolesnichenko, et al. 2020; Kobler, Sburlea, Mondini, et al. 2020).

# Conclusion

We find that our study (Ofner et al. 2017) was misinterpreted, and we have described why we think that the study protocol in (Bibián et al. 2021) is not suitable to assess the validity of our study. Furthermore, we have raised points that could explain why the multiclass classification in the constrained movement condition did not exceed chance level. We fully agree that artefacts are particularly a problem in low-frequency EEG signals as they often covary with the task, and paradigms need to be carefully designed with that in mind. We also would like to highlight in this context the importance of model interpretation (e.g., via EEG source imaging) for assessing whether signals were decoded from plausible brain regions.